\documentstyle[emulateapj,epsfig]{article}   

\slugcomment{Resubmitted to ApJ: February 11, 2002}
\lefthead{}

\begin{document}

\def\lmxb{neutron star}
\def\lmxbs{neutron stars}
\def\spose#1{\hbox to 0pt{#1\hss}}
\def\lta{\mathrel{\spose{\lower 3pt\hbox{$\mathchar"218$}}
     \raise 2.0pt\hbox{$\mathchar"13C$}}}
\def\gta{\mathrel{\spose{\lower 3pt\hbox{$\mathchar"218$}}
     \raise 2.0pt\hbox{$\mathchar"13E$}}}
\def\Msun{{\rm M}_\odot}
\def\msun{{\rm M}_\odot}
\def\Rsun{{\rm R}_\odot}
\def\Lsun{{\rm L}_\odot}
\def\half{{1\over2}}
\def\RL{R_{\rm L}}
\def\zs{\zeta_{s}}
\def\zR{\zeta_{\rm R}}
\def\dJJ{{\dot J\over J}}
\def\dMM{{\dot M_2\over M_2}}
\def\tKH{t_{\rm KH}}
\def\eck#1{\left\lbrack #1 \right\rbrack}
\def\rund#1{\left( #1 \right)}
\def\wave#1{\left\lbrace #1 \right\rbrace}
\def\dd{{\rm d}}

\title{A Unified Description of the Timing Features of Accreting 
       X-ray Binaries}
\author{Tomaso~Belloni}
\affil{Osservatorio Astronomico di Brera,
	Via E. Bianchi 46, I-23807 Merate (LC), Italy;
	belloni@merate.mi.astro.it}

\author{Dimitrios~Psaltis}
\affil{School of Natural Sciences,
       Institute of Advanced Study,\\ 
       Princeton, NJ 08540;
       dpsaltis@ias.edu}

\author{Michiel van der Klis}
\affil{Astronomical Institute ``Anton Pannekoek'', University of
	Amsterdam and Center for High-Energy Astrophysics,\\
        Kruislaan 403,NL 1098 SJ Amsterdam, the Netherlands;
	michiel@astro.uva.nl}

\def\rem#1{{\bf [ #1]}}\def\hide#1{}
\def\new#1{{ #1}}

\begin{abstract}
We study an empirical model for a unified description of the power
spectra of accreting neutron stars and black holes. This description
is based on a superposition of multiple Lorentzians and offers the
advantage that all QPO and noise components are dealt with in the same
way, without the need of deciding in advance the nature of each
component. This approach also allows us to compare frequencies of
features with high and low coherences in a consistent manner and
greatly facilitates comparison of power spectra across a wide range of
source types and states.  We apply the model to \new{six} sources, the
low-luminosity X-ray bursters 1E~1724$-$3045, SLX~1735$-$269 and 
GS~1826$-$24, the high-latitude transient XTE~J1118$+$480, 
\new{the bright system Cir X-1, and the Z source GX 17+2}. We
find that it provides a good description of the observed spectra,
without the need for a scale-free ($1/f$) component. We update
previously reported correlations between characteristic frequencies of
timing features in the light of this new approach and discuss
similarities between different types of systems which may point
towards similar underlying physics.
\end{abstract}


\section{INTRODUCTION}\label{intro}

Since the first detailed timing studies of accreting X-ray binaries
made with the {\em EXOSAT\/} and {\em Ginga\/} satellites, the
aperiodic- and quasi-periodic variability and spectral properties of
low-magnetic-field ($B\lesssim 10^{10}$~G) neutron stars (hereafter
simply neutron stars) and of accreting black-hole candidates (hereafter BHC)
have been considered in a rather different fashion, despite their
striking similarities.  In the case of \lmxbs, the sources were soon
classified into two separate classes (the ``Z'' and the ``atoll''
sources) based on their spectral and timing behavior (see van der Klis
1995a and references therein). This classification applied, with few
exceptions, to all bright systems; the fainter \lmxbs, although not
exhibiting the full complement of phenomenology, seemed to fit in as
well.  In the case of the BHCs, things appeared to be more complex and
from the beginning a classification as precise as in the previous case
appeared to be problematic (see Tanaka \& Lewin 1995; van der Klis
1995a). Despite this complexity, however, some common ground was found
between the phenomenologies of \lmxbs\ and BHCs (van der Klis 1994a,
1994b). In particular, strong similarities were pointed out in the
properties and dependence on X-ray luminosity and spectral shape of
the powerful broad band noise seen in both \lmxbs\ and BHCs, as well
as between some of the QPOs. These similarities were interpreted in
terms of particular source states occurring in both \lmxbs\ and BHCs
depending, presumably, on mass accretion rate.

With the advent of the Rossi X-ray Timing Explorer ({\em RXTE}), our
knowledge of the properties of the aperiodic variability of accreting
X-ray binaries took a substantial step forward. The power-spectral
densities of \lmxbs\ revealed two previously unknown QPOs with
frequencies in the kHz range which, in addition to broad continuum
components, led to power spectra containing up to five simultaneous
quasi-periodic oscillations (QPOs). A number of characteristic
frequencies could thus be identified for each power spectrum. The kHz
oscillations were soon found to behave in a rather regular way and
their phenomenological properties were fitted into the tight scheme
organizing the neutron-star systems (see van der Klis 2000 for a
review). BHCs showed fewer QPOs, superimposed on strong continuum
components, which provided a more limited number of characteristic
frequencies. Even though some new QPOs were discovered, such as the
``hecto-hertz'' QPOs (Morgan, Remillard \& Greiner 1997; Remillard et
al.\ 1999a, 1999b; Homan et al. 2001) and some $<$100~Hz QPOs turned
out to be clearly similar to those in \lmxbs, BHC phenomenology proved
once again more difficult to classify.  There are even indications
that the similarities between \lmxbs\ and BHCs disappear in the newly
accessible frequency range above $\sim$100 Hz (see, e.g., Sunyaev \&
Revnivtsev 2000).

Recently, two studies have revisited the above question and identified
a number of power-spectral components with frequencies that follow the
same correlations in both \lmxbs\ and BHCs. Wijnands \& van der Klis
(1999a, hereafter WK99) considered the break frequencies of the
broad-band noise and the centroid frequencies of the low-frequency
QPOs in the 0.02--70~Hz range and found these frequencies to follow a
similar correlation in \lmxbs\ and BHCs covering 2.5 decades in
frequency. Psaltis, Belloni, \& van der Klis (1999, hereafter PBK99)
systematically studied the numerous QPOs and peaked noise components
in the 0.1--1200~Hz range and identified two features whose
characteristic frequencies follow a correlation covering three orders
of magnitude in frequency. For bright \lmxb\ sources, these two
features are the ``horizontal branch oscillation'' (HBO) and the lower
kHz QPO, respectively. For BHCs and low-luminosity \lmxbs, they are
the the low-frequency QPO and a broad peaked component at frequencies
between 1 and 10 Hz.

These studies suggest that the same types of variability occur in both
neutron-star and black-hole sources, which can severely constrain
theoretical models for their variability properties.  Two major
classes of models exist to explain the variable-frequency QPOs in
\lmxbs: the beat frequency models (e.g., Alpar \& Shaham 1985;
Strohmayer et al.\ 1996; Miller, Lamb, \& Psaltis 1998), in which the
spin-frequency of the neutron star provides one of the characteristic
frequencies, and the relativistic precession models (e.g., Stella,
Vietri, \& Morsink 1999; Psaltis \& Norman 2001), in which all QPO
frequencies arise from general relativistic frequencies in the
accretion disk. The latter class of models do not depend explicitly on
the properties of the compact object and are therefore also applicable
to the case of the variable-frequency QPOs observed in BHCs.  At the
same time, discoseismic models offer a promising explanation for the
``hecto-Hertz'' QPOs in these sources (e.g., Nowak et al. 1997; Perez
et al.\ 1997; see Wagoner 1999 for a review).

Similarly to the various models for the {\em quasi-periodic\/}
variability properties of accreting compact objects, various attempts
have been made for understanding the physical origin of the broadband
{\em aperiodic} variability in both types of sources, which in fact
constitutes most of the signal in the power spectra of BHCs.  This
noise continuum can often roughly be described by an $1/f$ noise with
a characteristic low-frequency break and may, therefore, be produced
by a large variety of physical mechanisms.  Different models have
considered the contribution of time-variable inhomogeneities
(``blobs'') in the accretion disks (e.g., Bao and Ostgaard 1995;
Takeuchi, Mineshige, \& Negoro 1995), short-lived magnetic flares in
the accretion-disk coronae (Poutanen \& Fabian 1999), erratic
injections of soft photons in a Comptonizing medium (Kazanas \& Hua
1999), MHD instabilities in the accretion disks (Hawley \& Krolik
2001), etc.

From the early days of X-ray astronomy it was clear that the power
spectrum of the BHC Cyg X-1 is not a pure power law (Terrell 1972;
Nolan et al. 1981). With {\em EXOSAT\/} it became evident that the
power spectra of BHCs (Belloni \& Hasinger 1990a) as well as atoll
sources (Hasinger \& van der Klis 1989) are not smooth, but show
``bumps and wiggles''.  These have been variously described by
allowing breaks or exponential cutoffs in power-law fits, by adding
broad Lorentzian components, and combinations of these approaches
(see references above and, e.g., Berger and van der Klis 1998). For the
low-luminosity neutron stars, recent work by Olive et. al (1998) and
Barret et al. (2000) revealed power spectra that are quite similar to
those of BHCs and atoll sources in the low hard state. All these power
spectra are flat-topped, become steeper towards higher frequency and
show bumps and wiggles. An approach to the description of these power
spectra of \lmxbs\ and BHCs adopted by a number of authors (e.g.,
Miyamoto et al.\ 1991; Olive et al.\ 1998; Nowak 2000)
includes no power-law or broken power-law components in the fits
(except perhaps a single steep power law at the very lowest
frequencies), but uses only a sum of Lorentzian components, some of
which are broad (with coherence factor Q$\le$1). 
These authors demonstrated that the power
spectra can be well modeled without the need for power-law components;
the bumps and wiggles correspond to ``shoulders'' fitted with
individual Lorentzian components.

A limitation of the traditional approaches is that peaked components
(QPO) and broad-band components (noise) are treated a priori as different,
and therefore are fitted with intrinsically different models. This applies
to both BHC and NS systems.
In this paper, we expand and elaborate on the description of the
power spectra of accreting compact objects as the sum of a small
number of distinct Lorentzian components,
proposing it as a model to fit all timing features, independent of their
Q values. Here, we apply it to 
power density spectra of a sample of sources in specific states,
i.e. the Low/hard state of BHCs and the low-luminosity state of
X-ray bursters, but we plan to investigate its applicability on
a wider range of sources and states.
We seek to offer a unified
paradigm for the fitting and interpretation of the power-density
spectra of both accreting neutron stars and black holes. This will
help to unambiguously identify similar variability components in
neutron-star and black-hole sources and thereby to constrain models
for quasi-periodic and aperiodic variability of accreting compact
objects. We apply this method to \new{six} additional sources: the
low-luminosity X-ray bursters 1E~1724$-$3045, SLX~1735-269, and
GS~1826-24 (some of these data have already been presented by Olive et
al.\ 1998 and Barret et al.\ 2000), the recently discovered
high-latitude X-ray transient XTE~J1118$+$480, which is most probably
a BHC (McClintock et al. 2001; Wagner et al. 2001;
for the X-ray data see also Revnivtsev, Sunyaev
\& Borozdin 2000), \new{the bright and peculiar NS system Cir X-1 (see
Shirey et al. 1996), and the high-luminosity Z-source GX 17+2 (see 
Homan et al. 2002).}
Finally, we discuss the current status of the
correlations between timing features of accreting X-ray binaries in
the light of our new approach.

\section{A LORENTZIAN DECOMPOSITION OF THE OBSERVED POWER SPECTRA}
\label{decomp}

Following Nowak (2000), we describe the power spectra of both
neutron-star and black-hole sources by a sum of Lorentzian components,
given individually by
\begin{equation}
P(\nu) = {r^2\Delta\over \pi} {1\over \Delta^2 + (\nu-\nu_0)^2}\;,
\end{equation}
where $r$ is the integrated fractional rms (over $-\infty$ to
$+\infty$) of each Lorentzian and $\Delta$ its HWHM.  We also display
them as $\nu P_{\nu}$ vs.\ $\nu$ (see Belloni et al.\ 1997; Nowak
2000), where $P_{\nu}$ is the power density in rms normalization
(e.g., van der Klis 1995b) at frequency $\nu$ and quote as their
``characteristic frequency'' the frequency $\nu_{max}$, at which they

\centerline{
\epsfxsize=9.5truecm
\centerline{{\epsfbox{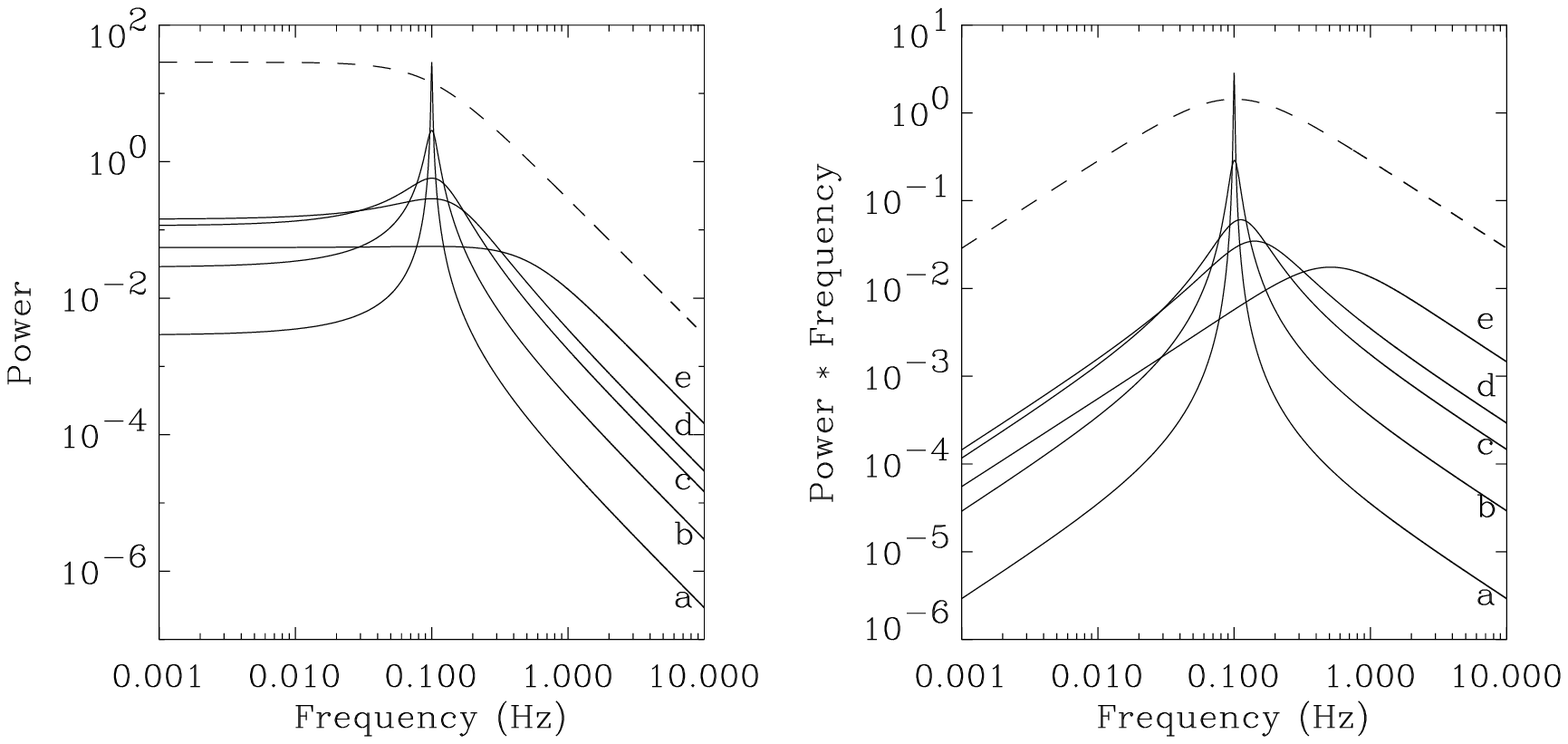}}}}
\figcaption[fig1.ps]{\footnotesize 
        Examples of Lorentzian models. Left panel: $P_\nu$
	representation; right panel: $\nu P_\nu$ representation.
	Solid lines represent Lorentzians with $\nu_0$=0.1 Hz
         and
	different $Q$ values: the spectra labeled a, b, c, d and e
	correspond to Q=50, 2, 1, 0.5 and 0.1, respectively. 
	The dashed line is a zero-centered Lorentzian with
	$\Delta$=0.1 Hz.}

\mbox{}

\noindent attain their maximum in $\nu P_\nu$, i.e.,
\begin{equation}
\nu_{max} = \sqrt{\nu_0^2 + \Delta^2} = \nu_0 \sqrt{1+{1/(4Q^2)}}
\end{equation}
(see Belloni et al.\ 1997), where $Q\equiv\nu_0/2\Delta$. This is the
frequency around which the component contributes most of its power per
logarithmic frequency interval and is a measure for the highest
frequency covered by the Lorentzian. Note that the maximum in $\nu
P_\nu$ does {\it not} correspond to the centroid frequency $\nu_0$ of
the Lorentzian but $\nu_{max} \ge \nu_0$.  The difference is small for
narrow QPOs, but becomes large in the case of broad features.

In Figure~1 we show a Lorentzian with $\nu_0$=0.1~Hz and different
values of the coherence $Q$. The left panel shows the usual $P_\nu$
vs.\ $\nu$ representation whereas the right panel shows $\nu P_\nu$
vs.\ $\nu$.  The shift in $\nu_{max}$ as Q decreases is evident. As
the half-width $\Delta$ increases at constant $\nu_0$ in order to
produce a broader feature, the value of $\Delta$ progressively takes
over from $\nu_0$ in determining the characteristic frequency
$\nu_{max}$. The logarithmic FWHM in $\nu P_\nu$ of the feature
asymptotically approaches an upper limit of
$\log{2+\sqrt3\over2-\sqrt3}$.  For very low values of $Q$, the
function approaches a zero-centered Lorentzian ($Q$=0), which in $\nu
P_\nu$ peaks at $\Delta$. For comparison, the dashed line in Fig.~1
corresponds to a zero-centered Lorentzian with $\Delta$=0.1 Hz.

For fitting very broad features, even those with a centroid frequency
that is formally larger than zero, a zero-centered Lorentzian model is
often adequate. As a matter of fact, this is what a fitting procedure
does: if a power spectral component is narrow ($Q\gg$1), such a
routine converges to a value of $\nu_0$ corresponding to the centroid
of the observed peak, setting $\Delta$ equal to its half-width. If, on
the other hand, the component is wide ($Q<$1), the routine chooses a
value of $\nu_0$ approaching 0 and set $\Delta$ to fit the
characteristic frequency of the feature (see Fig.~1). Of course, if
the component is too broad, one never obtains a good fit (this would
be the case if the power spectra were pure power laws).  We discuss
below how to deal with the intermediate cases.

Traditionally, the term QPO has been restricted to features with $Q>$2
(see van der Klis 1995a), while less coherent components have been
classified as `noise'. Broad peaked components are sometimes called
`peaked noise'.  As noted by van der Klis (1995b) this classification
is arbitrary; there is a continuous spectrum of intermediate cases,
and indeed, there is evidence that a component can sometimes change
from being a clear band-limited noise component into a clear QPO or vice
versa (Di Salvo et al.\ 2000). Noise components can represent
characteristic variability frequencies just as QPOs do. However, there
is an ambiguity in how characteristic frequencies are usually reported
for narrow (i.e., QPOs) and broad (i.e., noise) variability
components. For QPOs, two parameters are usually quoted, i.e., the
centroid frequency $\nu_0$ and the FWHM in $P_\nu$, but it is $\nu_0$
which is usually called the ``characteristic frequency'' of the
component. For noise components, only one parameter is usually quoted,
for example one describing the width of the component, such as the
HWHM $\Delta$, or some power-law break frequency
$\nu_{break}$. (Simulations show that fitting zero-centered
Lorentzians with broken power laws and vice-versa yields similar
values for the $\Delta$ of the Lorentzian and the $\nu_{break}$ of the
broken power-law, but with systematic differences, see
Section~\ref{secwk99}.)  So, in intermediate cases, the definition of
the characteristic frequency is ambiguous; it depends on whether the
component is considered to be a QPO or noise.

Based on the discussion above, we adopt the following algorithm for
describing the power spectra. (1) We fit a model that is the sum of
$n$ Lorentzians and involves no power-law or broken power-law
components.  (2) As the characteristic frequency of each component we
adopt its peak frequency in the $\nu P_\nu$ spectrum, i.e.,
$\nu_{max}\equiv\sqrt{\nu_0^2+\Delta^2}$. As noted, if a component is
narrow, its characteristic frequency $\nu_{\rm max}$ corresponds to
the central frequency $\nu_0$ of the Lorentzian (see eq.~[2]), whereas,
if it is broad, the characteristic frequency approaches $\Delta$.  For
very low-Q components it makes little difference to either the derived
characteristic frequency or the fit to the data to set $\nu_0\equiv0$.

\begin{deluxetable}{rrrcr}
\scriptsize
\tablecolumns{5}
\tablecaption{Log of the observations.\label{observ}}
\tablewidth{0pt}
\tablehead{}
\startdata
      &\multicolumn{2}{c}{Date} & Number of     & Exposure \\
Obs.  & Start    &  End         & power spectra & time (s) \\
\hline
\multicolumn{5}{c}{XTE J1118+480} \\
\hline
A & 00-04-13 &	00-05-04 & 243 & 31104 \\
B & 00-05-06 &	00-05-15 & 136 & 17408 \\
\hline
\multicolumn{5}{c}{1E 1724-3045} \\
\hline
C & 96-11-04 &	96-11-08 & 239 & 30592 \\
D & 97-02-25 &	97-03-25 &  70 &  8960 \\
E & 97-04-18 &	97-05-20 &  59 &  7552 \\
F & 98-02-11 &	98-02-15 & 236 & 30208 \\
G & 98-02-19 &	98-02-20 & 383 & 49024 \\
H & 98-02-26 &	98-02-28 & 243 & 31104 \\
I & 98-03-16 &	98-03-18 & 147 & 18816 \\
J & 98-04-08 &	98-04-10 & 232 & 29696 \\
K & 98-05-30 &	98-05-31 & 119 & 15232 \\
L & 98-06-04 &	98-06-04 &  95 & 12160 \\
\hline
\multicolumn{5}{c}{SLX 1735-269} \\
\hline
M & 97-10-10 &	97-10-12 & 254 & 32512 \\
\hline
\multicolumn{5}{c}{GS 1826-24} \\
\hline
N & 97-11-05 &	97-11-06 & 348 & 44544 \\
O & 98-06-07 &	98-06-08 & 144 & 18432 \\
P & 98-06-12 &	98-06-12 & 102 & 13056 \\
Q & 98-06-23 &	98-06-23 &  54 &  6912 \\
R & 98-06-24 &	98-06-25 &  87 & 11136 \\
\hline
\multicolumn{5}{c}{GX 17+2} \\
\hline
S & \multicolumn{4}{c}{see Homan et al. (2002)}\\
\hline
\multicolumn{5}{c}{Cir X-1} \\
\hline
T & 96-03-12 &  96-03-12 &  32 &  2048 \\
U & 96-03-16 &  96-03-16 &  19 &  1216 \\
\tableline
\enddata
\end{deluxetable}

Of course, in the standard ``lifetime broadening'' interpretation of a
Lorentzian as the power spectrum of an exponentially damped harmonic
oscillator, the Lorentzian centroid and width correspond to {\it two}
different physical time scales (respectively, the oscillation period
and damping time), whereas in the corresponding interpretation of a
zero-centered Lorentzian the width is a measure for the exponential
decay time of an exponential shot. Our description of both these types
of components in terms of a single characteristic frequency
$\nu_{max}$ is motivated primarily by empirical considerations and has
no immediate interpretation within the damped oscillator picture. We
note, however, that in interpretations where broad power spectral
components arise by the superposition of variations covering a {\it
range} of frequencies rather than by lifetime broadening, $\nu_{max}$
is just a measure for the highest frequency represented in that
range, \new{or the one which contributes the highest power.} 
Identifying, in an accretion disk model, frequencies with radii
in the disk would lead to a link between $\nu_{max}$ and the inner
radius of the disk annulus contributing to the component.
\new{More insight on the connection to a physical model might come
from the comparison of the fractional rms values of the different
components in addition to their characteristic frequencies. As
discussed in Sect. 5, two important observational facts are that
the relative contribution of the broad components in low-frequency
systems is such as to produce an approximate power law distribution, 
and that the highest frequency component in BH systems is considerably
suppressed with respect to NS systems.}

Our choice of components and parameters poses the problem of
comparison with previous results. In this paper, we apply conversions
from $\nu_0$ and $\Delta$ into $\nu_{max}$ only for the data from
Nowak (2000). All other data discussed when dealing with global
correlations will have to be examined for complete consistency, but
since most of them involve kHz QPOs, which have a rather high Q, it
follows that $\nu_{max}\approx\nu_0$, so the effects are small. The
Lorentzian components of the two power spectra in Nowak (2000) with
reasonably constrained parameters have $Q$ values ranging from 0.6 to
2.0, corresponding to correction for their characteristic frequencies
of 1.9 to 1.03 respectively (see eq.~[2]). As will be discussed below,
we believe that fitting all components with Lorentzians will
eventually facilitate comparison across different types of sources, as
it allows us to treat broad noise components and QPOs in formally the
same way without having to decide in advance how to interpret each
component. This makes it easier to compare components in different
sources and allows us to make the connection with the fits to power
spectra of neutron-star sources with much higher characteristic
frequencies.

\section{OBSERVATIONS}\label{obs}

In order to test our approach on power spectra containing complex
broad-band components,
we analyzed RXTE/PCA observations of one black-hole candidate and
\new{five} neutron-star systems, \new{covering a wide range of characteristic
frequencies}. The black-hole candidate is the recently
discovered X-ray transient XTE~J1118$+$480 (Remillard et al. 2000,
Revnivtsev, Sunyaev \& Borozdin 2000). This source was in a low hard
state during these observations The neutron-star systems are three
low-luminosity X-ray bursters (see Barret et al. 2000): 1E~1724$-$3045
(see also Olive et al. 1998), SLX~1735$-$269, and GS~1826$-$24, 
\new{the peculiar
system Cir X-1 (see Shirey et al. 1996), and the bright Z-source GX~17+2
(see Homan et al. 2002).}
\new{The three bursters} are usually considered to 
be atoll sources which are always in the low (island) state.  The
observation log is reported in Table~1.

For each observation, usually consisting of several continuous
intervals corresponding to separate RXTE orbits, we produced a series
of power spectra from data segments lasting \new{64 or} 128 seconds
\new{(depending on the total length of the observation)} and a time
binning of 1/1024 seconds, normalized after Leahy et al. (1983).
These spectra were then averaged together and logarithmically
rebinned. The contribution of the Poissonian statistics was removed
(see van der Klis 1995a) and the spectra were converted to squared
fractional rms (Belloni \& Hasinger 1990b; van der Klis 1995b). The
average power spectra from different observations were compared and,
if they were compatible, they were further averaged together.  The
total resulting numbers of power spectra averaged can be found in
Table 1. Notice that our dataset B for XTE~1118$+$480 is the same as
that analyzed by Revnivtsev, Sunyaev, \& Borozdin (2000), and that our
dataset C for 1E~1724$-$3045 corresponds to the data presented by Olive
et al. (1998). Also, the power spectra from datasets M (SLX~1735$-$269)
and N (GS~1826$-$24) have been presented in Barret et al. (2000). To
each of our final average power spectra we fitted a combination of
Lorentzians as described in the previous section. For the fits we used
XSPEC V11.0.19. The only exception to the procedure outlined above is the 

\centerline{
\epsfxsize=9.5truecm
\centerline{{\epsfbox{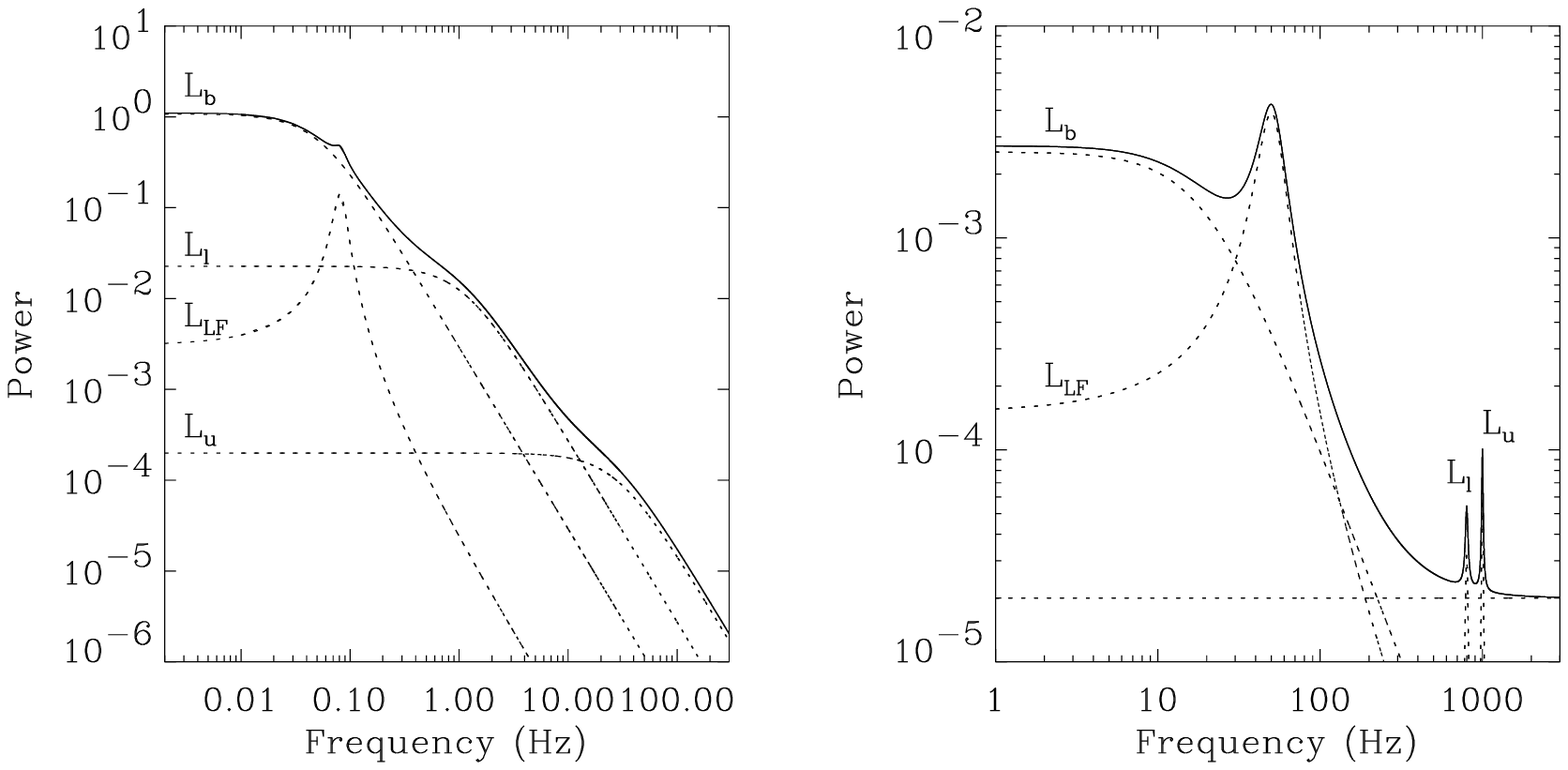}}}}

\figcaption[fig2.ps]{\footnotesize 
 		Left panel: model decomposition for XTE~1118$+$480.
 		Right panel: the same four Lorentzian components but
 		with characteristic frequencies corresponding to the
 		high-frequency end of the WK99 and PBK99 correlations
 		and narrower widths for the two high-frequency
 		components.  }
\mbox{}

\noindent dataset for GX 17+2, for which we used one of the power 
density spectra from Homan et al. (2002) for the fits.

In practice, it turned out that all power spectra \new{(with the exception 
of that for GX 17+2, see below)} could be fitted with
the following components (see Fig. 2, left panel):

\begin{itemize}

\item A zero-centered low-frequency Lorentzian L$_b$ fitting the 
low-frequency end (the flat top) of the band-limited noise visible in
all power spectra, with characteristic (``break'') frequency $\nu_b$.

\item Two zero-centered Lorentzians L$_\ell$ and L$_u$ fitting the 
high-frequency end of the band-limited noise, with (``lower'' and
``upper'') characteristic frequencies $\nu_\ell$ and $\nu_u$.

\item One or two Lorentzians fitting the region around the
low-frequency QPO. The profile of this QPO is sometimes more complex
than a simple Lorentzian, consisting of a relatively narrow core
L$_{LF}$ (the ``LF QPO'') at characteristic frequency $\nu_{LF}$, and
a broader ``hump'' component L$_h$ with characteristic frequency
$\nu_h$.

\end{itemize}

It is evident that, in this description, the band-limited noise is
fitted with three broad (and in our case always zero-centered)
Lorentzians. For the lowest-frequency Lorentzian, a flat top is
actually seen in the power spectra, while the true behavior at low
frequencies of the other two components is masked by components
that are more powerful there. The lowest-frequency Lorentzian, L$_b$,
produces the break at $\nu_b$ in the band-limited noise, the
highest-frequency one, L$_u$, produces the second break at $\nu_u$,
and the middle one, L$_\ell$, essentially ``fills the hole'' between
the two, so that between $\nu_b$ and $\nu_u$ a relatively smooth,
approximately $1/f$, spectrum is created, with some ``bumps and
wiggles'', as described by, e.g., Hasinger and van der Klis (1989) and
Belloni and Hasinger (1990a), and on which the LF QPO is
superimposed. We found that always
$\nu_b<\nu_{LF}\lta\nu_h<\nu_\ell<\nu_u$; when both the hump L$_h$ and
the LF QPO L$_{LF}$ are present, their {\it centroid} frequencies $\nu_h^0$
and $\nu_{LF}^0$ typically coincide, but as the hump is broader, usually
$\nu_{LF}<\nu_h$. The three components together describing the noise,
L$_b$, L$_\ell$ and L$_u$, were always required by the data, although
$\nu_u$ was not in all cases well-constrained. Of the two components
used to describe the low-frequency QPO region, L$_{LF}$ and L$_h$, at
least one was always required.  Notice that the frequencies for
L$_{LF}$ and L$_h$ presented in Tables 2 and 3 are centroid
frequencies and HWHM, which are the values obtained by fitting the
data: for these two components the characteristic frequencies
discussed in the text have subsequently been computed by applying
Equation 2. 

\new{For the case of GX 17+2, which has much higher characteristic frequencies,
the main differences with respect to the scheme described above are: (a) the
L$_b$ component is very broad but not zero-centered; 
(b) the L$_{LF}$ component
shows two additional peaks corresponding to a sub-harmonic and a first
overtone (see Homan et al. 2002); 
(c) the two higher-frequency components L$_\ell$ and L$_u$ are in this case
narrow kHz QPO peaks, as expected; 
(d) two more broad components are needed at frequencies
below $\nu_u$.}

\begin{deluxetable}{ccccccc}
\scriptsize
\tablecolumns{7}
\tablecaption{Best-fit characteristic frequencies
	      for XTE~1118+480. For L$_{LF}$ centroid frequencies are quoted.
		\label{1118}}
\tablewidth{0pt}
\tablehead{}
\startdata
& L$_b$&\multicolumn{2}{c}{L$_{LF}$}& L$_\ell$ & L$_u$ \\
\hline
Obs.& $\nu_b$ (Hz)    & $\nu^0_{LF}$ (Hz)   & $\Delta$ (Hz) & $\nu_\ell$ (Hz) & $\nu_u$ (Hz) & $\chi^2$(d.o.f.)\\
\hline
A   & 0.062$\pm$0.004 & 0.115$\pm$0.003&0.009$\pm$0.004&1.27$\pm$0.04 &25.88$\pm$3.61 & 97.4(73)\\
B   & 0.052$\pm$0.002 & 0.081$\pm$0.002&0.006$\pm$0.003&1.10$\pm$0.03 &27.88$\pm$3.09 & 156.9(73)\\
\tableline
\enddata
\end{deluxetable}

\subsection{XTE J1118$+$480}

The power spectra of XTE~J1118+480 was fitted with four components
(see Fig.~3), the LF QPO already reported by Revnivtsev, Sunyaev \&
Borozdin (2000), and the three zero-centered Lorentzians L$_b$,
L$_\ell$, and L$_u$ for the band-limited noise.  The best-fit
parameters are reported in Table~2. All components are significant.
The values of $\chi^2$ are high, but an examination of the residuals
shows no systematic trend. 
The $Q$ value for the LF QPO is around 7.

\vspace{0.5cm}

\epsfxsize=8.truecm
\centerline{{\epsfbox{1118.ps}}}
\figcaption[1118.ps]{\footnotesize 
	Power spectra in $\nu P_\nu$ form for XTE~J1118+480. Lines
	mark the best-fit model and its components.}
\mbox{}

\epsfxsize=8.truecm
\centerline{{\epsfbox{terzan2.ps}}}
\figcaption[terzan2.ps]{\footnotesize
	Power spectra in $\nu P_\nu$ form for 1E~1724-3045
	(Obs.\ H \& D). Lines mark the best-fit model and its
	components.}

\mbox{}

\subsection{1E~1724$-$3045}

To fit the ten power spectra of 1E~1724$-$3045, five Lorentzian
components are necessary. In the first three power spectra (C,D,E),
the component in the LF QPO region is rather broad, with a $Q$ of
0.7--0.8 (the same component can be seen in Olive et al.\ 1998).  In
the remaining power spectra, an additional narrow peak appears, with
$Q>$1, sometimes as high as $Q\simeq 15$.  By simple comparison of the
spectra, we identify the narrow component as the LF QPO L$_{LF}$, and
the broad component as the ``hump'' L$_h$. In Figure~4 two
representative power spectra are shown: in the power spectrum of
observation H the narrow LF QPO component L$_{LF}$ is clearly visible
while in D it is absent.  Note that in the $\nu P_\nu$ representation
the hump peaks at a higher frequency than the corresponding 

\epsfxsize=8.truecm
\centerline{{\epsfbox{slx1735.ps}}}
\figcaption[slx1735.ps]{\footnotesize
	Power spectrum in $\nu P_\nu$ 
	form for SLX~1735-269 (Obs. M).
	Lines mark the best-fit model and its
        components.}

\mbox{}

\noindent narrow LF QPO. The centroid frequencies of these two 
components are the same, so in a $P_\nu$ plot their peaks coincide.
For observation C the narrow LF QPO is marginally present, but not
required by the fit, while for observation E, like D, it is not
detected (see Table~3).

\subsection{SLX~1735$-$269}

For SLX~1735-269, we produced only one power spectrum, using the same
data as Barret et al.\ (2000). There are additional datasets in the
RXTE public archive, but their total exposure time is not sufficient
for the generation of good power spectra. The power spectrum is shown
in Figure~5.  A four-Lorentzian model was used to fit the data. The best
fit parameters are reported in Table~3.  As noted by Barret et
al.\ (2000), they are remarkably similar to those of some observations
of 1E~1724$-$3045.  Since there is no narrow component, we identify the
second Lorentzian as the ``hump'' component L$_h$.

\begin{deluxetable}{ccccccccc}
\scriptsize
\tablecolumns{9}
\tablecaption{Best-fit characteristic frequencies
	    for 1E 1724-3045, SLX~1735-269 and
		GS~1826-24. For L$_{LF}$ and L$_h$ centroid
frequencies are quoted. \label{Ter2}}
\tablewidth{0pt}
\tablehead{}
\startdata
& L$_b$&\multicolumn{2}{c}{L$_{LF}$}&\multicolumn{2}{c}{L$_h$}& L$_\ell$ & L$_u$ \\
\hline
Obs.& $\nu_b$ (Hz)    & $\nu_{LF}^0$ (Hz)   & $\Delta$ (Hz) & $\nu_h^0$ (Hz)	& $\Delta$ (Hz) & $\nu_\ell$ (Hz) & $\nu_u$ (Hz) & $\chi^2$(d.o.f.)\\
\hline
\multicolumn{9}{c}{1E~1724-3045} \\
\hline
C   & 0.195$\pm$0.010 &---&---& 0.87$\pm$0.03&0.64$\pm$0.03&10.64$\pm$0.51 &191.20$\pm$29.30 & 166.0(73)\\
D   & 0.626$\pm$0.038 &---&---& 2.91$\pm$0.13&1.81$\pm$0.29&13.50$\pm$3.80 &256.60$\pm$64.50 & 62.0(70)\\
E   & 0.331$\pm$0.026 &---&---& 1.59$\pm$0.09&1.15$\pm$0.12&14.99$\pm$2.14 &198.70$\pm$53.70 & 90.0(70)\\
F   & 0.150$\pm$0.008 &0.63$\pm$0.03&$<$0.06& 0.62$\pm$0.02&0.57$\pm$0.02& 9.91$\pm$0.37 &200.30$\pm$27.25 & 108.6(67)\\
G   & 0.165$\pm$0.012 &0.69$\pm$0.03&$<$0.19& 0.72$\pm$0.06&0.62$\pm$0.05& 9.70$\pm$0.52 &166.70$\pm$27.18 & 97.1(67)\\
H   & 0.124$\pm$0.006 &0.50$\pm$0.03&$<$0.05& 0.51$\pm$0.01&0.49$\pm$0.02& 8.05$\pm$0.20 &142.40$\pm$11.88 & 141.6(67)\\
I   & 0.170$\pm$0.009 &0.69$\pm$0.02&$<$0.02& 0.70$\pm$0.03&0.61$\pm$0.03& 9.46$\pm$0.52 &203.20$\pm$34.89 & 93.9(67)\\
J   & 0.180$\pm$0.008 &0.78$\pm$0.02&$<$0.33& 0.79$\pm$0.03&0.66$\pm$0.03&10.72$\pm$0.44 &208.50$\pm$27.98 & 119.1(67)\\
K   & 0.221$\pm$0.017 &1.01$\pm$0.02&$<$0.10& 1.03$\pm$0.07&0.87$\pm$0.07&11.66$\pm$1.01 &184.10$\pm$39.70 & 70.3(67)\\
L   & 0.193$\pm$0.014 &0.86$\pm$0.02&$<$0.30& 0.85$\pm$0.07&0.79$\pm$0.07&11.77$\pm$0.79 &178.50$\pm$32.53 & 108.0(67)\\
\hline
\multicolumn{9}{c}{SLX 1735-269} \\
\hline
M   & 0.179$\pm$0.014 &---&---& 0.87$\pm$0.11&0.78$\pm$0.11&10.40$\pm$1.63 & 74.65$\pm$33.77 &	94.4(73)\\
\hline
\multicolumn{9}{c}{GS~1826-24} \\
\hline
N   & 0.272$\pm$0.006 &1.13$\pm$0.02&$<$0.07& 1.37$\pm$0.03&1.07$\pm$0.03&15.07$\pm$0.39 &353.80$\pm$46.98 & 122.1(70)\\
O   & 0.273$\pm$0.009 &1.00$\pm$0.18&$<$0.10& 1.22$\pm$0.04&1.07$\pm$0.04&13.54$\pm$0.55 &256.40$\pm$46.23 & 128.9(70)\\
P   & 0.441$\pm$0.017 &1.82$\pm$0.03&$<$2.22& 2.05$\pm$0.06&1.44$\pm$0.06&20.09$\pm$1.47 &456.25$\pm$82.88 & 72.8(70)\\
Q   & 0.901$\pm$0.056 & ----	    &---    & 4.26$\pm$0.18&2.13$\pm$0.35&22.20$\pm$4.93 &$>$593.85   & 85.0(73)\\
R   & 0.604$\pm$0.038 &2.28$\pm$0.07&$<$0.45& 1.54$\pm$0.15&1.58$\pm$0.25&18.61$\pm$2.21 &$>$662.40   & 106.7(70)\\
\tableline
\enddata
\end{deluxetable}

\subsection{GS~1826$-$24}

We obtained five power spectra of GS~1826$-$24. The first one
(observation N) corresponds to that presented by Barret et al.\
(2000). The two X-ray bursts present in the data of observation N
(Barret et al.\ 2000) as well as four more X-ray bursts present in the
remaining datasets were excluded from the analysis. The ``hump''
component was present in all spectra; in four of the power spectra a
narrow LF QPO was also observed. Notice that in this source there are
significant differences between the centroid frequencies of L$_h$ and
L$_{LF}$. In observations Q and R only 

\epsfxsize=8.truecm
\centerline{{\epsfbox{gs1826.ps}}}
\figcaption[gs1826.ps]{\footnotesize
	Power spectra in $\nu P_\nu$ form for GS~1826-24 (Obs. N \& Q).
	Lines mark the best-fit model and its components.}

\mbox{}

\noindent a lower limit could be obtained on the characteristic ``upper''
frequency $\nu_u$.  Figure~6 shows the two power spectra of
GS~1826$-$24 with the highest and lowest characteristic
frequencies. Notice the narrow LF QPO component present in the power
spectrum of observation N and absent from observation Q.  The best fit
parameters can be found in Table~3.

\subsection{Cir X-1}

A detailed description of the power density spectra of Cir X-1 can be
found in Shirey et al. (1996, 1998). The L$_{LF}$ component in this
source was seen to vary from 1 to $\sim$12 Hz (see also PBK99). We
selected two observations among those reported by Shirey et
al. (1996), one at relatively low ($\sim$1.3 Hz) and one at relatively
high ($\sim$8.5 Hz) characteristic frequencies.  In order to avoid
effects due to the variability of the source, for observation U we
selected only the last RXTE orbit, corresponding to the lowest
characteristic frequencies.  Both power density specta can be fitted
with three Lorentzian components (see Table 4 and Fig. 7)): a
zero-centered one (L$_b$), a narrow one (L$_{LF}$), and a broad one at
high frequencies (L$_\ell$).  

\begin{deluxetable}{ccccccccc}
\scriptsize
\tablecolumns{9}
\tablecaption{Best-fit characteristic frequencies
            for GX 17+2 and Cir X-1. All frequencies are $\nu_{max}$.
            The sub-harmonic and first harmonic of L$_{LF}$ are not
	    reported here. (a): values from Homan et al. (2002), not
	    from our fits.
            \label{GXCir}}
\tablewidth{0pt}
\tablehead{}
\startdata
Obs.&
$\nu_{b'}$ (Hz)  &
$\nu_{b''}$ (Hz)  &
$\nu_b$ (Hz)     & 
$\nu_{LF}$ (Hz)  & 
$\nu_\ell$ (Hz)  & 
$\nu_u$ (Hz)     & 
$\chi^2$(d.o.f.) \\
\hline
\multicolumn{7}{c}{GX 17+2} \\
\hline
S   & 
0.87$\pm$0.11  &
3.05$\pm$0.29  &
6.93$\pm$1.05  &
38.10$\pm$0.25  &
537$\pm$23$^a$  &
798$\pm$6$^a$  &
150.2(138)\\

\hline
\multicolumn{7}{c}{Cir X-1} \\
\hline
T   & 
---            &
---            &
0.59$\pm$0.03   &
1.27$\pm$0.02   &
24.09$\pm$3.75  &
---           &
169.6(151)\\

U   & 
---            &
---            &
4.77$\pm$0.16   &
8.43$\pm$0.04   &
109.22$\pm$10.09 &
---           &
153.5(168)\\

\tableline
\enddata
\end{deluxetable}

\subsection{GX 17+2}

In order to include one example of a NS source with high
characteristic frequencies (and kHz QPOs), we selected one observation
of the Z-source GX 17+2 from the sample analyzed by Homan et
al. (2002), corresponding to $S_z$ between 0.5 and 0.6 (middle of the
horizontal branch, see Homan et al. 2002).  Notice that, since the two
kHz peaks in this observation are quite well separated from the other
components, our fits below 200 Hz do not affect the kHz QPOs, so that
for these high-frequency peaks we could adopt the values reported in
Homan et al. (2002).  We obtained the best fit with a model consisting
of six Lorentzian components (see Fig. 8): a broad ($Q$=0.6) component
which we identify with L$_b$, a narrow one with a sub-harmonic and a
first harmonic (see Fig. 8) identified with L$_{LF}$, and two
additional broad ($Q<1$) components with characteristic frequencies
below $\nu_b$, which we indicate as L$_b'$ and L$_b''$. Notice that
the L$_b'$ component is not very significant (about 3$\sigma$). The
best fit parameters can 

\epsfxsize=8.truecm
\centerline{{\epsfbox{cirx1.ps}}}
\figcaption[gx17.ps]{\footnotesize
	Power spectrua in $\nu P_\nu$ form for Cir~X-1 (Obs.\ T \& U).
        Lines mark the best-fit model and its components.}

\mbox{}
 
\noindent be found in Table 4.

\subsection{Summary of the fits and comparison with power-law models}

The three zero-centered Lorentzians L$_b$, L$_\ell$ and L$_u$ are
required in \new{many} spectra to fit the band-limited noise. A narrow
low-frequency QPO L$_{LF}$ is detected in the black hole candidate,
and in {\it some} of the neutron star spectra (preferentially those
with the lower characteristic frequencies -- note that the frequencies
are low in the BHC as well). The broad hump component is required in
\new{most} neutron-star spectra, but not in the BHC. In 1E~1724$-$3045, when
L$_{LF}$ and L$_h$ are simultaneously present, their centroid

\epsfxsize=8.truecm
\centerline{{\epsfbox{gx17.ps}}}

\figcaption[gx17.ps]{\footnotesize 
	Power spectrum in $\nu P_\nu$ form for GX 17+2 (Obs. S).
        Lines mark the best-fit model and its components. Notice that, unlike
	all other cases, this power spectrum is not rms-normalized.}

\mbox{}

\noindent frequencies are identical, but in GS~1826$-$24 these frequencies are
usually different.

In order to compare the quality of our current fits with that of other
models \new{for broad components}, we also fitted the power spectra
first with a combination of a broken power law and one Lorentzian for
the narrow LF QPO peak. This model has 7 free parameters (compared
with the 9 of our model). We applied this model to two representative
power spectra (observations C of 1E~1724$-$3045 and A of
XTE~J1118$+$480). The best fit $\chi^2$ values were 1220 (/75 dof) and
478 (/75 dof) respectively, to be compared with our values 166(/73
dof) and 156(/73 dof). Clearly, the multi-Lorentzian model fits much
better.  A model consisting of a twice-broken power law (Belloni and
Hasinger 1990a) plus a LF QPO, with 9 free parameters, gives similar
$\chi^2$ values as our multi-Lorentzian fit. As noted in
\S\ref{decomp}, the advantage of our current approach is that it
facilitates comparison across source types.

\section{GLOBAL CORRELATIONS}\label{global}

As PBK99 and Nowak (2000) noted, the correlations observed between
\new{the frequencies of the} LF QPO and both kHz QPOs in
luminous neutron stars can be extended to lower frequencies by
relations between the LF QPO and two
broad Lorentzians respectively taking over the role of the lower kHz QPO 
(seen in several low-luminosity neutron stars and in BHCs) and
the upper kHz QPO (seen in Cyg X-1 and GX 339$-$4).
\new{Our results} may fit in with this. The frequencies $\nu_b$, 
$\nu_h$ or $\nu_{LF}$, $\nu_\ell$, and $\nu_u$ \new{define a} model
that is formally similar to that for luminous neutron stars.  In
Figure~2 the left panel, the four frequencies are those of our fit to
the BHC XTE~1118$+$480. In the right panel, they represent a power
spectrum of a luminous neutron-star Z source (e.g., van der Klis et
al.\ 1997); here the high-frequency components have been made
narrower.  The band-limited noise L$_b$ and the low-frequency QPO
$L_{LF}$ are both common also at high luminosities (van der Klis
1995a). A broad peaked noise component (``bump'' or ``wiggle'')
similar to our L$_\ell$ component is seen in several higher luminosity
sources as well (e.g., Shirey et al.\ 1998; PBK99) and may be a
low-frequency manifestation of the lower kHz QPO (PBK99). Finally,
L$_u$ was first found by Nowak (2000) in Cyg~X-1 and GX~339$-$4 and
tentatively identified as a low-frequency version of the upper kHz
QPO. The presence of such a component was already noted previously
(but not interpreted in the same way) by Olive et al.\ (1998) for
1E~1724$-$3045 (our observation C; see also Yoshida et al.\ 1993).

The power spectra can be more complex than the sum of four Lorentzians
(see, e.g., Ford and van der Klis 1998; Di Salvo et al.\ 2001; van
Straaten et al.\ 2000 and our GX 17+2 power spectrum).  In particular,
the low-frequency QPO often shows multiple harmonics, the second
sometimes being the strongest one (e.g., Belloni et al. 1997). In the
following we use the frequencies of the strongest QPO peaks in the
power spectrum \new{(but see Sect. 4.2)}.  We try to identify the four
components discussed here in all source classes, and investigate if
their frequencies follow or extend the correlations of WK99 and
PBK99. If these components at very different frequencies are thus
phenomenologically related, perhaps they are physically related. In
addition to our results, we include additional new points on Cyg~X-1
and GX~339$-$4 (Nowak 2000), XTE~J1550$-$564 (Homan et al.\ 2001),
4U~1915$-$05 (Boirin et al.\ 2000), and 4U~1728$-$34 (Di Salvo et al.\
2001).

\subsection{Correlation between the flat-top power level and the break
frequency}

In Cyg X-1 the break frequency $\nu_b$ and the power level of the flat
part of the band-limited noise P$_{\rm flat}$ are anti-correlated
(Belloni \& Hasinger 1990). From their figure, it is evident that
P$_{\rm flat}\propto 1/\nu_b$.  The high-frequency end of the power
spectrum remains approximately unchanged. This is seen in other BHCs
(Yoshida et al. 1993; van der Klis 1994a; M\'endez \& van der Klis
1997), in the neutron star 4U~0614+09 (van Straaten et al. 2000), and
also clearly in our two sources showing significant changes in their
characteristic frequencies (Fig.~9 and Fig.~10).

The Lorentzian description of the band-limited noise suggests a clue
to this effect. From Equation~(1), for a zero-centered Lorentzian,
$P_{\rm flat} = r^2/\pi \Delta$, i.e., when the fractional rms $r$ of
the Lorentzian is constant, $P_{\rm flat}$ is proportional to
$1/\Delta$, and hence, as $\nu_0=0$, to $1/\nu_{max}$. Hence, the
observed effect can be reproduced by allowing the characteristic
frequency of L$_b$ to vary, while keeping its total fractional rms
constant.  The dotted lines in Fig.~10 represent the predicted $P_{\rm
flat}$ vs.\ $\nu_b$ relation for constant $r$. The definition of $r$
used here (Eq.~1) involves integration from 

\centerline{
\epsfxsize=8.truecm
\centerline{{\epsfbox{bheffect.ps}}}}
\figcaption[bheffect.ps]{\footnotesize
	Power spectra of 1E~1724-3045 (top panel) and GS~1826-24
	(bottom panel) from this work. They are overlaid for
	comparison of their relative fractional rms and characteristic
	frequencies.}

\mbox{}

\noindent $-\infty$ to $+\infty$, so
the true values of the rms are a factor $\sqrt{2}$ less.  Most points
approximately follow such correlations, with a difference between
sources in normalization (i.e., different $r$).

So, the total rms of L$_b$ is rather constant in each of our sources
but different between sources. The neutron stars all cluster near an
rms of $\sim$20\%, whereas the two BHCs have values of $\sim$30\% and
$\sim$40\%, respectively, consistent with what would be expected if
rms correlates with a basic compact-object property such as mass. The
lower rms in neutron-star sources could also result from the
contribution of a non-varying component from the stellar surface.

\subsection{Correlation between the low-frequency QPO and break
frequency} \label{secwk99}

This correlation (WK99) contains data from both BHCs and \lmxbs, to
which we add our data except SLX 1735$-$269 (which does not have a LF
QPO, see Table 3), the two accurate points from Table~1 in Nowak
(2000), GX~339-4 (full dataset) and Cyg~X-1 (4-15 keV data), corrected
for the low-Q shift (see Sect. 2). \new{and Q-corrected GX 339-4
points from Nowak, Wilms \& Dove (2002), for which we plot their Q2
component as our l$_h$}. For Cyg X-1 we used the first ``harmonic'' of
the peak, which is by 

\centerline{
\epsfxsize=8.truecm
\centerline{{\epsfbox{bhall.ps}}}}
\figcaption[bhall.ps]{\footnotesize
		Correlation between the break frequency $\nu_b$ and
		flat-top level of the corresponding zero-centered Lorentzian.
		Filled circles are from Belloni \& Hasinger (1990), other
		symbols are from this work. The dotted lines represent
		lines of constant total rms 
		for a Lorentzian model. Numbers along the right-hand
axis are $r^2$ as defined in Equation~1, i.e., a factor of $\sqrt2$ larger
than the true noise amplitude.}

\mbox{}

\noindent far the strongest (see Nowak 2000 and Sect. 4).
We also included data for 4U~1728-34 (Di Salvo et al. 2001), but not
for 4U~1915-05 (Boirin et al.\ 2000) or XTE~J1550-564 (Homan et al.\
2001), as these authors report parameters for the QPOs only.

The result is shown in Figure~11. The two frequencies follow the same
trend for over three orders of magnitude, but the individual
correlations in different sources are significantly different. Various
systematic effects affect the data.  The characteristic frequency of
the break depends on energy and hence on source spectrum and energy
band (Belloni et al.\ 1997). WK99 used broken power laws, while we
used zero-centered Lorentzians.  By simulations, we find that this
causes over- or underestimates of the break frequency by $\lesssim
15$\%, unlikely to explain the largest of the offsets we observe.
Some discrepancies are consistent with factors of 2 or 4, suggesting
that that some of the QPO frequencies are not fundamental. Indeed, the
sub-harmonic in GX 17+2 lies on the WK99 correlation.

Some sources show an additional broad component, L$_h$, at a slightly
higher characteristic frequency than L$_{LF}$ (see Sect. 3), which
might represent the non-Lorentzian wings of the narrow LF QPO.  Its
characteristic frequency $\nu_h$ is well correlated with $\nu_{LF}$:
the corresponding points are added in Figure~11.  This correspondence
between a narrow QPO and a broader component at a similar centroid
frequency may be related to the cases in which a narrow QPO appears
near the break of a band-limited noise (e.g., Morgan, Remillard, \&
Greiner 1997 in GRS 1915+105; Di Salvo et al.\ 2001 in 4U 1728$-$34;
Wijnands \& van der Klis

\centerline{
\epsfxsize=9.truecm
\centerline{{\epsfbox{wk.ps}}}}
\figcaption[wk.ps]{Correlations between characteristic frequencies
$\nu_{LF}$ and $\nu_b$ of variability in \lmxbs\ and BHCs. Small
points are from WK99, large symbols are from this work and from Nowak
(2000) and Di Salvo et al. (2000). The crosses below the main relation
around 1 Hz show the correlation between L$_h$ and L$_{LF}$. }

\mbox{}

\noindent 1999b and Revnivtsev, Borozdin \& Emelyanov
1999 in XTE J1806--246; Homan et al. 2001 in XTE J1550--564).
\new{Notice that this L$_h$ component is clearly present in the 
GX 339-4 data from Nowak, Wilms, \& Dove (2002): when only one component
is present, it is L$_h$, when two (or more) are present, they are L$_{LF}$ 
and L$_h$ (see Fig. 11).}

\subsection{Correlation between low-frequency QPO and L$_\ell$}

Figure 12 shows the correlation of PBK99, together with the Nowak
(2000), Boirin et al.\ (2000), Homan et al.\ (2001) Di Salvo et al.\
(2001) data. \new{and Nowak, Wilms, \& Dove (2002) data.  Most points
agree quite well with the correlation, except two GS~1826$-$24 points
and the Homan et al.  (2001) points (see below).}

The Nowak (2000) points would be further above the correlation without
Q-correction. \new{Most of the GX 339-4 points by Nowak, Wilms \& Dove
(2002) lie a factor of 2 above the main correlation, again an
indication that the frequency adopted for L$_{LF}$ might not be that
of the fundamental.}  For the QPO frequencies of XTE J1550-564 (Homan
et al.  2001) we used the strongest of the low-frequency harmonic
peaks for L$_{LF}$ and the 185$-$285 Hz QPOs for L$_\ell$.  These
points do not follow the main correlation of PBK99, but extend its
``second branch'', but points corresponding to the beginning of the
outburst (Cui et al. 1999) in the original PBK99 correlation, follow
the main branch.  The points of 4U~1728$-$34 by Di Salvo et al. (2001)
and of 4U~1915$-$05 by Boirin et al. (2000) also follow partly the
main branch and partly the second branch.

Di Salvo et al. (2001) concluded that the feature plotted on this
second branch vs. $\nu_\ell$ is in fact, in our current terminology,
$L_b$, and hence the main and second branch in PBK99 together
represent the WK99 relation, in 

\centerline{
\epsfxsize=9.truecm
\centerline{{\epsfbox{pbk_low.ps}}}}
\figcaption[pbk.ps]{\footnotesize
	$\nu_{LF}$ and $\nu_\ell$ of variability in \lmxbs and
	BHCs. Small points are from PBK99, large symbols are from this
	work and from Nowak (2000), Di Salvo et al. (2000), Boirin et
	al. (2000), Barret et al. (2000).}

\mbox{}

\noindent accordance with their observation that
L$_b$ changes from band-limited noise into a QPO when its frequency
increases. Indeed, if we plot $\nu_b$ vs. $\nu_\ell$, the points
cluster around the extension of the second branch to lower frequencies
(not shown in Fig. 12).

\subsection{Correlation between the frequencies of components L$_\ell$ and L$_u$}

To this correlation (Fig. 13) between upper and lower kHz QPO, we
added new data for 4U~1728$-$34 (Di Salvo et al.\ 2001) and 4U~1915-05
(Boirin et al. 2000), as well as our $L_u$ components and those by
Nowak (2000).
Nowak's BHC points (with Q-correction) and SLX~1735$-$269 follow the
low-frequency extension of the correlation, but the BHC
XTE~J1118$+$480 and the neutron stars 1E~1724$-$3045 and GS~1826$-$24
are above it.  Perhaps this discrepancy is related to the presence of
other components in the power spectrum in the few 100 Hz range.  A
broad Lorentzian with an aprozimately constant $100-200$~Hz frequency
occurs in low-luminosity \lmxbs\ (see, e.g., Ford \& van der Klis
1998; van Straaten et al. 2000; Di Salvo et al.\ 2001) and might also
occur in these sources (van Straaten et al. 2001).

\section{DISCUSSION}\label{disc}

Applying the model described in Sect. 2, we have shown that the power
spectra of low-luminosity burst sources, of a BHC in the low state,
\new{of a Z source, and of Cir X-1} can be decomposed into the sum of
a small number ($4-5$) of Lorentzian components.  We obtain, for each
power spectrum, four characteristic frequencies: $\nu_b$, $\nu_{LF}$
or $\nu_h$, $\nu_\ell$, and $\nu_u$, which include the broad noise
components as well as the narrow QPOs identified in previous studies.
In this manner, the identification of similar power spectral
components between different sources (as done in PBK99) becomes easier
and free of potentially subjective interpretation (see, however, Nowak
2000 and \S\ref{decomp} for a discussion

\centerline{
\epsfxsize=9.truecm
\centerline{{\epsfbox{pbk_high.ps}}}}
\figcaption[pbk.ps]{\footnotesize
        Correlations between characteristic frequencies
        $\nu_\ell$ and $\nu_u$ of variability in \lmxbs\  and
        BHCs. Small points are from PBK99, large symbols are from this
        work and from Nowak (2000), Di Salvo et al. (2000), Boirin et
        al. (2000), Barret et al. (2000).  }

\mbox{}

\noindent  of some complications related
to the wings and harmonic structure of the low-frequency QPO).

The presence in some power spectra of the additional L$_h$ component
needs to be discussed in some more detail. This component seems to be
closely related to L$_{LF}$. Sometimes only one of these two
components is observed; sometimes both are present. When they appear
simultaneously, their centroid frequencies are approximately the same
in 1E~1724$-$3045 but not in GS~1826$-$24.  As already mentioned,
there are cases of power spectra in the literature (see, e.g., Morgan,
Remillard, \& Greiner 1997), for which the characteristic frequency of
a band-limited noise component is rather close to that of a
low-frequency QPO. Moreover, Di Salvo et al.\ (2001) actually observed
the transformation of one into the other, which also involved
'transitory' power spectra where both a narrow and a broad component
seem to be present simultaneously, with the QPO sitting on the break
of the noise spectrum. It seems likely that the connection between
$\nu_h$ and $\nu_{LF}$ is a similar case. Based on their frequency
coincidence, the two components may be related physically as
well. However, unlike Di Salvo et al.\ (2001), we observe that, in the
case discussed here, there is a preference for the presence of the
narrow LF QPO when the characteristic frequencies are {\it low}, not
when they are high.  That the narrow LF QPO component tends to
disappear in the spectra with the highest characteristic frequencies
suggests that caution is necessary when identifying this LF QPO with
those seen at higher luminosities. It is possible that, after one
narrow Lorentzian disappears, another one comes into prominence, as
observed in 4U~1728-34 (di Salvo et al. 2001).  A more detailed
discussion of the characteristics of this component in a number of
BHCs will be presented in a forthcoming paper.

We have also studied the correlations between the properties of the
power-spectral components and extended the correlations found earlier
for a wide range of other accreting sources. Our results are
consistent with those of earlier studies, but also indicate some
discrepancies, which may arise from the fact that different sources
may follow similar but not identical correlations (see also
Psaltis et al.\ 1998 where this was demonstrated quantitatively
for the case of kHz QPOs in bright neutron-star sources).  Before
discussing in detail the nature of these discrepancies, a more
homogeneous analysis using identical models would be necessary
for all sources. 

Our description of the power spectra in terms of just Lorentzian
components characterized by a single characteristic frequency
$\nu_{max}$ puts noise and QPO phenomena on equal footing. This
approach is in accordance with recent findings from the analysis of
{\em RXTE\/} data of the atoll source 4U~1728$-$34 by Di Salvo et al.\
(2001), who showed that in some cases a broad flat-top noise component
turns into a narrow, peaked component, therefore becoming a
``standard'' QPO. The fact that $\nu_{max}$ of components of widely
differing $Q$ seems to maintain (at least on some occasions) a simple
relation to other parameters suggests that perhaps the power
distribution of these components is related to the extent of disk
annuli whose inner radius, setting $\nu_{max}$, varies in a simpler
fashion than their width (\S\ref{decomp}). 
It might also be related
to the presence of a lifetime broadening mechanism that operates at a
characteristic timescale related to the dynamical frequencies in the
disk, such as viscous dissipation (see, e.g., Psaltis \& Norman
2001). Finally, in a shot-noise model of the broad-band variability,
large changes in the $Q$ value of a component can be produced by a
change in the correlation time between different shots (see, e.g.,
Vikhlinin, Churazov \& Gilfanov\ 1994), accounting thus for such changes.
However, as noted in Sect. 2, life-time broadening mechanisms do not naturally
involve $\nu_{max}$ as a basic parameter; rather, centroid and width reflect
aspects of the physical process which are a priori independent.

Independent of the particular physical mechanism that is responsible
for the various power-spectral components, our results strongly
suggest that the low-frequency Lorentzian (or flat-topped power-law
component) is not different from the other components. Indeed,
not only can narrow QPOs turn into flat-topped components, as
discussed above, but also the frequency $\nu_{\rm b}$ is correlated to
the frequencies of all narrow QPOs. The latter property, together with
the fact that in BHCs the low-frequency ``noise'' accounts for almost
the entire power in X rays, suggests that this component
is generated at the same region in the accretion flow as the other
QPOs (see, e.g., Psaltis \& Norman 2001).

Our analysis, with the help of the improved statistics in the {\em
RXTE/PCA\/} data as compared to previous, less sensitive missions
suggests that no $1/f$ noise component is needed for the
interpretation of power spectra in BHCs and low-luminosity bursters.
However, the superposition of Lorentzian components does produce an
overall power spectrum that in neutron stars resembles an $1/f$
distribution over a range of frequencies, and in the BHC XTE J1118+480
is clearly steeper, but also roughly follows a power law. The relative
total rms strengths of the various Lorentzians are such that a
relatively smooth power spectrum (in the $P_\nu$ vs.\ $\nu$
representation) is produced. This property needs to be addressed by
theoretical models as much as the values for the characteristic
frequencies do.

Within the context of our results, we can also discuss the recent
suggestion by Sunyaev \& Revnivtsev (2000) of a possible way of
distinguishing between systems hosting a black hole and a
weakly-magnetic neutron star based on their power spectra.  According
to these authors, neutron-star systems can show significant power
above $\sim 300$~Hz, while BHCs do not. The relevant effect can be
seen comparing Figure~3 to Figure~4: the power spectrum of
1E~1724-3045 contains much more power at high frequencies than that of
XTE~1118+480, as was already noted by Sunyaev \& Revnivtsev (2000) and
Revnivtsev, Sunyaev \& Borozdin (2000) for these two sources
respectively.  

Our power-spectral decomposition suggests that the difference in the
power spectra of BHCs and \lmxbs\ is caused by two effects. The first
is that all characteristic frequencies are lower in the case of the
BHC, probably reflecting the mass dependence of the dynamical time
scale in the accretion flow, as was also noted by Sunyaev \&
Revnivtsev (2000). The second is more unexpected: in the power
spectrum of XTE~1118$+$480 the fractional rms of the Lorentzian with
the highest frequency is considerably lower with respect to the other
components. This effect is not easy to understand. However, the
fact that the component is significantly detected in this BHC, albeit
weaker than in neutron stars, does seem to suggest that it is not a
feature that {\it requires} a neutron-star surface for its generation;
apparently a component with these characteristics can also be produced
by the accretion disk around a black hole.

In conclusion, we suggest that the underlying physical mechanisms
producing similar components in all these sources may be the same and
that the scatter in their different correlations is due to a
combination of differences in the analysis method and differences in
the fundamental system parameters. Although not {\it all} timing
features in black-hole and neutron-star low-mass X-ray binaries are
included in this representation (e.g., the VLFN, the normal branch
oscillations in Z and perhaps some atoll sources---but see PBK---, the
$\simeq 1$~Hz QPO in dipping \lmxbs, the $\simeq 100$~Hz QPOs in atoll
sources---but see \S 4.4---, and the HS power-law noise as well as the
very-low frequency QPOs in some BHCs), the ones included here are
important and their understanding would be a major step forward in the
study of variability in accreting X-ray sources.

\acknowledgements
We thank S.\ Campana, L.\ Stella, and C.\ Dullemond for useful
discussions, D.\ Barret for help with importing power spectra into
XSPEC, and J. Homan for having provided the power density spectrum of
GX 17+2. T.\,B.\ acknowledges the hospitality of
Harvard-Smithsonian CfA and of MIT, and thanks the Cariplo Foundation for
financial support.  D.\,P.\ has been supported by a post-doctoral
fellowship of the Smithsonian Institution and by the NASA Long Term
Space Astrophysics program under grant NAG~5-9184. 
This work was supported in part by the Netherlands Organization for 
Scientific Research (NWO).


\clearpage

\begin{thebibliography}{}

\bibitem[Alpar \& Shaham 1985]{as85}
	Alpar, M.A., Shaham, J., 1985, Nature, 316, 239

\bibitem[Bao \& Ostgaard 1995]{bo95}
	Bao, G, \& Ostgaard, E.\ 1995, \apj, 443, 54

\bibitem[Barret et al. 2000]{bar00}
	Barret, D., Olive, J.F., Boirin, L., Done, C., Skinner, G.K.,
	Grindlay, J.E., 2000, ApJ, 533, 329

\bibitem[Belloni \& Hasinger 1990a]{bh90a}
	Belloni, T., Hasinger, G., 1990a, A\&A, 227, L33

\bibitem[Belloni \& Hasinger 1990b]{bh90b}
	Belloni, T., Hasinger, G., 1990b, A\&A, 230, 103

\bibitem[Belloni et al. 1997]{bel97}
	Belloni, T., van der Klis, M., Lewin, W.H.G. van Paradijs, J.,
	Dotani, T., Mitsuda, K., Miyamoto, S., 1997, A\&A, 322, 857

\bibitem[Berger \& van der Klis 1991]{bk91}
	Berger, M., \& van der Klis, M.\ 1998, A\&A, 340, 143

\bibitem[Boirin et al. 2000]{boi00}
	Boirin, L., Barret, D., Olive, J.F., Bloser, P.F., Grindlay, J.E.,
	2000, A\&A, 361, 121

\bibitem[Churazov et al. 2001]{cgr01}
	Churazov, E., Gilfanov, M., Revnivtsev, M., 2001, MNRAS, 321, 759

\bibitem[Cui et al. 1999]{cui99}
	Cui, W., Zhang, S.N., Chen, W., Morgan, E.H., 1999, ApJ, 512, L43

\bibitem[Di Salvo et al. 2001]{sal01}
	Di Salvo, T., M\'endez, M., van der Klis, M., Ford, E.,
	Robba, N.R., 2001, ApJ, 546, 1107

\bibitem[Ford \& van der Klis 1998]{hk1998}
	Ford, E.C., van der Klis, M., 1998, ApJ, 506, L39

\bibitem[Hasinger \& van der Klis 1989]{hk1989}
	Hasinger, G., \& van der Klis, M., 1989, A\&A, 225, 79

\bibitem[Hawley \& Krolik 2001]{hk2001}
	Hawley, J.\,F., \& Krolik, J.\,H.\ 2001, \apj, 548, 348

\bibitem[Homan et al. 2001]{hom01}
	Homan, J., Wijnands, R., van der Klis, M., Belloni, T., van Paradijs,
	J., Klein-Wolt, M., Fender, R.P., M\'endez, M., 2001, ApJSuppl,
        132, 377

\bibitem[Homan et al. 2002]{hom02}
        Homan, J., van der Klis, M., Jonker, P.G., Wijnands, R., Kuulkers, E.,
	M\' endez, M., Lewin, W.H.G., 2002, Apj, in press (astro-ph/0104323)

\bibitem[Kazanas \& Hua 1999]{kh99}
	Kazanas, D., \& Hua, X.-M. 1999, \apj, 519, 750

\bibitem[Leahy et al. 1983]{lea83}
	Leahy, D. A., Darbro, W., Elsner, R. F.,
	Weisskopf, M. C., Kahn, S.,
	Sutherland, P. G., Grindlay, J. E., 1983, ApJ, 266, 160

\bibitem[McClintock et al. 2001]{mc01}
	McClintock, J.E., Garcia, M.R., Caldwell, N., Falco, E.E., 
	Garnavich, P.M., Zhao, P., 2001, ApJ, 511, L147

\bibitem[M\'endez \& van der Klis 1997]{mk97}
	M\'endez, M., van der Klis, M., 1997, ApJ, 479, 926

\bibitem[Miyamoto et al.\ 1991]{m91}
	Miyamoto, S., Kimura, K., Kitamoto, S., Dotani, T., \& 
	Ebisawa, K.\ 1991, \apj, 383, 784

\bibitem[Miller et al. 1998]{mil98}
	Miller, M.C., Lamb, F.K., Psaltis, D., 1998, ApJ, 508, 791

\bibitem[Morgan et al. 1997]{mrg97}
	Morgan, E.H., Remillard, R.A., Greiner, J., 1997, ApJ, 482, 993

\bibitem[Nolan et al. 1981]{nol81}
        Nolan, P.L., Gruber, D.E., Matteson, J.L., Peterson, L.E., 
	Rothschild, R.E., Doty, J.P., Levine, A.M., Lewin, W.H.G.,
	Primini, F.A., 1981, ApJ, 246, 494

\bibitem[Nowak et al. 1997]{n97}
	Nowak, M.A., Wagoner, R.V., Begelman, M.C.,
	Lehr, D.E., 1997, ApJ, 477, L91

\bibitem[Nowak 2000]{n00}
	Nowak, M.A., 2000, MNRAS, 318, 361

\bibitem[Nowak, Wilms \& Dove 2002]{n02}
        Nowak, M.A., Wilms, J., Dove, J.B., 2002, MNRAS, in press
        (astro-ph/0201383)

\bibitem[Olive et al. 1998]{ol98}
	Olive, J.F., Barret, D., Boirin, L., Grindlay, J.E., Swank, J.H.,
	Smale, A.P., 1998, A\&A, 333, 942

\bibitem[Perez et al. 1997]{Petal97} 
	Perez, C.\,A., Silbergleit, A.\,S., Wagoner, B.\,V., \& Lehr, 
	D.\,E.\ 1997, \apj, 476, 589

\bibitem[Poutanen \& Fabian 1991]{pf91}
	Poutanen, J., \& Fabian, A.\,C.\ 1999, \mnras, 306, L31

\bibitem[Psaltis et al. 1998]{ps98}
	Psaltis, D., M\'endez, M., Wijnands, R., Homan, J., Jonker, P.G.,
	van der Klis, M., Lamb, F.K., Kuulkers, E., van Paradijs, J.,
	Lewin, W.H.G., ApJ, 501, L95

\bibitem[Psaltis et al. 1999]{pbk99}
	Psaltis, D., Belloni, T., van der Klis, M., 1999, ApJ, 520, 262 (PBK99)

\bibitem[Psaltis \& Norman 2001]{pn01}
	Psaltis, D., Norman, C., 2001, ApJ, in press (astro-ph/0001391)

\bibitem[Remillard et al. 1999a]{rem99a}
	Remillard, R.A., Morgan, E.H., McClintock, J.E., Bailyn, C.D.,
	Orosz, J.A., 1999a, ApJ, 522, 397

\bibitem[Remillard et al. 1999b]{rem99b}
	Remillard, R.A., McClintock, J.E., Sobczak, G.J., Bailyn, C.D.,
	Orosz, J.A., Morgan, E.H., Levine, A.M., 1999b, ApJ, 517, L127

\bibitem[Remillard et al. 2000]{rem00}
	Remillard, R.A., Morgan, E.H., Smith, D., Smith, E., 2000,
	IAU Circ., 7389
 
\bibitem[Revnivtsev et al. 1999]{rbe99}
	Revnivtsev, M., Borozdin, K., Emelyanov, A., 1999, A\&A, 344, L25

\bibitem[Revnivtsev et al. 2000]{rsb00}
	Revnivtsev, M., Sunyaev, R., Borozdin, K., 2000, A\&A, 361, L37

\bibitem[Shirey et al. 1996]{shi96}
	Shirey, R.E., Bradt, H.V., Levine, A.M., Morgan, E.H.,
	1996, ApJ, 469, L21

\bibitem[Shirey et al. 1998]{shi98}
	Shirey, R.E., Bradt, H.V., Levine, A.M., Morgan, E.H.,
	1998, ApJ, 506, 374

\bibitem[Stella et al. 1999]{svm99}
	Stella, L., Vietri, M., Morsink, S.M., 1999, ApJ, 524, L63

\bibitem[Strohmayer et al. 1996]{str96}
	Strohmayer, T.E., Zhang, W., Swank, J.H., Smale, A., Titarchuk, L.,
	Day, C., Lee, U., 1996, ApJ, 469, L9

\bibitem[Sunyaev \& Revnivtsev 2000]{sr00}
	Sunyaev, R., \& Revnivtsev, M.\ 2000, A\&A, 358, 617

\bibitem[Takeuchi et al. 1995]{TMN95}
	Takeuchi, M., Mineshige, S., \& Negoro, H.\ 1995, \pasj, 47, 617

\bibitem[Tanaka \& Lewin 1995]{tl95}
	 Tanaka, Y., \& Lewin, W.H.G., 1995, in ``X-ray binaries'',
	 eds. Lewin, W.H.G., Van Paradijs, J., \& Van den Heuvel,
	 E.P.J., Cambridge Univ. Press., Cambridge, p2.

\bibitem[Terrell 1972]{te72}
	Terrell, N.J. Jr., 1972, ApJ, 174, L35

\bibitem[van der Klis 1994a]{vdk94a}
	 van der Klis, M., 1994a, A\&A, 283, 469

\bibitem[van der Klis 1994b]{vdk94b}
	 van der Klis, M., 1994b, ApJSuppl., 92, 511

\bibitem[van der Klis 1995a]{vdk95a}
	 van der Klis, M., 1995a, in ``X-ray binaries'',
	 eds. Lewin, W.H.G., Van Paradijs, J., and van den Heuvel,
	 E.P.J., Cambridge Univ. Press., Cambridge, p252.

\bibitem[van der Klis 1995b]{vdk95b}
	 van der Klis, M., 1995b, in ``The lives of Neutron Stars',,
	eds. Alpar, M.A., K\i z\i lo\u{g}lu, \"U, and van Paradijs, J.,
	NATO ASI 450, p301.

\bibitem[van der Klis 2000]{vdk00}
	van der Klis, M., 2000, Ann. Rev. Astr. Ap., 38, 717

\bibitem[van der Klis et al. 1997]{vdk97}
	van der Klis, M., Wijnands, R., Horne, K., Chen, W., 1997,
	ApJ, 481, L97

\bibitem[van Straaten et al. 2000]{vs00}
	van Straaten, S., Ford, E.C., van der Klis, M., M\'endez, M., 
	Kaaret, P., 2000, ApJ, 540, 1049

\bibitem[van Straaten et al. 2001]{vs01}
	van Straaten S., van der Klis, M., Di Salvo, T., Belloni, T., 
	Psaltis, D., 2001, ApJ, in press

\bibitem[Vikhlinin et al. 1994]{vk94}
	Vikhlinin, A., Churazov, E., Gilfanov, M., A\&A, 287, 73

\bibitem[Wagner et al. 2001]{W2001}
	Wagner, R.M., Foltz, C.B., Shahbaz, T., Casares, J., Charles, P.A.,
	Starrfield, S.G., Hewett, P., 2001, ApJ, 556, 42

\bibitem[Wagoner 1999]{W1999}
	 Wagoner, R.\,W.\ 1999, Phys.\ Rep., 311, 259

\bibitem[Wijnands \& van der Klis 1999a]{wk99a}
	Wijnands, R., van der Klis, M., 1999a, ApJ, 514, 939 (WK99)

\bibitem[Wijnands \& van der Klis 1999b]{wk99b}
	Wijnands, R., van der Klis, M., 1999b, ApJ, 522, 965

\bibitem[Yoshida et al. 1993]{yos93}
	Yoshida, K., Mitsuda, K., Ebisawa, K., Ueda, Y., Fujimoto, R.,
	Yaqoob, T., Done C., 1993, PASJ, 45, 605

\end{thebibliography}
\end{document}